\documentclass{ws-procs9x6}

\begin{document}

\title{SUPERCONDUCTIVITY AND ELECTRIC FIELDS:\\
A RELATIVISTIC EXTENSION OF\\
BCS SUPERCONDUCTIVITY}

\author{JAN GOVAERTS$^{1,2}$ and
DAMIEN BERTRAND$^{1,3}$}

\address{
$^{1}$Center for Particle Physics and Phenomenology (CP3),\\
Institute of Nuclear Physics, Catholic University of Louvain,\\
2, Chemin du Cyclotron, B-1348 Louvain-la-Neuve, Belgium\\
E-mail: Jan.Govaerts@fynu.ucl.ac.be\\
\vspace{7pt}
$^{2}$International Chair in Mathematical Physics and Applications (ICMPA),\\
University of Abomey-Calavi,\\
072 B.P. 50, Cotonou, Republic of Benin\\
\vspace{7pt}
$^{3}$Center for Space Radiations (CSR),\\
Department of Physics, Catholic University of Louvain,\\
2, Chemin du Cyclotron, B-1348 Louvain-la-Neuve, Belgium\\
E-mail: bertrand@spaceradiations.be}

\begin{abstract}
The effects of static electric fields on the superconducting
state are studied within a relativistic extension of the BCS theory
of superconductivity.
\end{abstract}

\vspace{40pt}

\begin{center}
To appear in the Proceedings of the Fourth International Workshop on Contemporary
Problems in Mathematical Physics,\\
November 5$^{\rm th}$-11$^{\rm th}$, 2005, Cotonou (Republic of Benin),\\
eds. J. Govaerts, M. N. Hounkonnou and A. Z. Msezane\\
(World Scientific, Singapore, 2006).
\end{center}

\vspace{40pt}

\begin{flushright}
CP3-06-11\\
ICMPA-MPA/2006/28
\end{flushright}

\bodymatter

\clearpage

\section{Introduction}
\label{Sec1}

The Bardeen--Cooper--Schrieffer (BCS) theory\cite{BCS}
of 1957 provides a microscopic understanding for the phenomena of 
Low Temperature (LTc) superconductivity.\cite{Tinkham,Waldram} 
Below the critical temperature $T_c$, attractive electron-phonon interactions
lead to electron-electron Cooper pair formation in the $s$-wave channel. 
The ensuing Cooper pair condensation 
of such identical bosonic quantum states implies the appearance of 
an energy gap, associated to a spontaneous breaking of the U(1) 
local gauge symmety of the electromagnetic interaction,\cite{Weinberg}
hence also an effective non-zero mass for the photon
which translates into the physical Meissner effect\cite{Meissner}
of magnetic field
screening in any bulk superconductor. The existence of a gap 
$\Delta(\vec{r}\,)$ also ensures the phenomenon of perfect conductivity,
through the collective dynamics of the condensed Cooper pair electrons
for electric currents less than some critical value.

The gap $\Delta(\vec{r}\,)$ may also be given the interpretation,
up to normalisation, of the common complex valued quantum wave function 
of the spin 0\, Cooper pairs. It also plays the role of an order parameter
for the phase transition towards the superconducting state, of
relevance in an effective field theory description. Among the
successes of the BCS theory in the weak coupling regime, 
one finds the correct description of the temperature dependence of 
the order parameter, hence the identification of the critical temperature, 
the critical magnetic field for the Meissner effect, and consequently 
also the critical current, inclusive of subtle effects such as 
the isotopic dependence of the critical temperature.

In effect, the superconducting state is understood in terms of a
coherent superposition of electron-electron pairs of which the
momentum and spin values are coupled in order to build up a
spin 0 state of vanishing total momentum, in the absence of
any electric current,
\begin{equation}
\prod_{\vec{k}}\left[u(\vec{k}\,)\,+\,
e^{i\theta(\vec{k}\,)}v(\vec{k}\,)\,
c^\dagger_{\downarrow}(-\vec{k}\,)
c^\dagger_{\uparrow}(\vec{k}\,)\right],
\label{eq:Cooper}
\end{equation}
where $c^\dagger_{\uparrow}(\vec{k}\,)$ and 
$c^\dagger_{\downarrow}(\vec{k}\,)$
represent the creation operators of electron states of momentum $\vec{k}$
and spin projection up or down, respectively, while $u(\vec{k}\,)$ and
$v(\vec{k}\,)$ stand for the probability amplitudes of occupation of
states without or with a single Cooper pair of vanishing total
momentum and spin. These two functions are identified through a gap
equation expressing the minimisation of the energy of such a trial
state with respect to these two functions obeying a normalisation condition
involving the combination $|u(\vec{k}\,)|^2+|v(\vec{k}\,)|^2$.

\clearpage

A few years after the formulation of the BCS theory, 
Gor'kov showed\cite{Gorkov} how through a finite temperature quantum 
field theory approach, it is possible to construct an effective field 
theory representation of the microscopic dynamics, which for 
all practical purposes coincides with the famous Ginzburg--Landau (GL) 
phenomenological description of superconductivity dating back 
to 1950 already.\cite{GL} Based on Landau's
approach towards a general theory of phase transitions, within
the GL theory the free energy density of the superconducting state, $F_s$,
compared to that of the normal state, $F_n$, is expressed as
a functional of the order parameter $\psi(\vec{r}\,)$ which, up
to normalisation, is identified with the superconducting gap 
$\Delta(\vec{r}\,)$,
\begin{equation}
F_s-F_n=\alpha|\psi|^2+\frac{1}{2}\beta|\psi|^4+
\frac{\hbar^2}{2m}\left|\left(\vec{\nabla}-i\frac{q}{\hbar}\vec{A}\right)\psi\right|^2
+\frac{1}{2\mu_0}\left(\vec{B}-\vec{B}_{\rm ext}\right)^2\ ,
\end{equation}
where $\alpha$ and $\beta$ are temperature dependent coefficients defining
an effective potential energy density
\begin{equation}
V\left(|\psi|\right)=\alpha|\psi|^2+\frac{1}{2}\beta|\psi|^4\ ,
\label{eq:Higgs}
\end{equation}
while the notation for the other quantities is standard, and corresponds
to the magnetic vector potential, $\vec{A}$, the magnetic induction,
$\vec{B}$, and the externally applied magnetic induction, $\vec{B}_{\rm ext}$,
with $\mu_0$ being the vacuum magnetic permittivity.
Finally, $q=-2|e|$ and $m$ stand, respectively, for the Cooper electric
charge and effective mass in the conducting material. Given the above
potential energy, as soon as the parameter $\alpha$ turns negative below
a specific critical temperature $T_c$, $\alpha(T<T_c)<0$, one has a potential of the Higgs
type with minima attained for nonvanishing expectation values of $\psi$,
thereby spontaneously breaking the local U(1) phase invariance symmetry of the
GL functional.

As explained in any standard textbook on 
superconductivity,\cite{Tinkham,Waldram} the phenomena
of perfect conductivity and diamagnetism are readily established from the
GL equations for the order parameter, namely the variational equations
of motion stemming from the GL functional,
\begin{equation}
\begin{array}{rl}
-\frac{\hbar^2}{2m} &
\left(\vec{\nabla}-i\frac{q}{\hbar}\vec{A}\right)^2\psi(\vec{r}\,)\,+\,
\alpha\,\psi(\vec{r}\,)\,+\,\frac{1}{2}\beta\,|\psi(\vec{r}\,)|^2\,
\psi(\vec{r}\,)\,=0,\\
 & \\
\vec{J}(\vec{r}\,)&=\ \frac{1}{\mu_0}\vec{\nabla}\times\vec{B}(\vec{r}\,) \\
 & \\
&=\ -i\frac{q\hbar}{2m}\left(\psi^*(\vec{r}\,)\vec{\nabla}\psi(\vec{r}\,)\,-\,
\vec{\nabla}\psi^*(\vec{r}\,)\,\psi(\vec{r}\,)\right)\,-\,
\frac{q^2}{m}\,|\psi(\vec{r}\,)|^2\,\vec{A}(\vec{r}\,),
\end{array}
\label{eq:GL}
\end{equation}
with boundary conditions requiring that the current $\vec{J}(\vec{r}\,)$ has
a vanishing component normal to any boundary in the case of a finite domain.
In particular, space dependence of the order parameter may then be 
accounted for, so that not only is the Meissner effect characterized 
by the magnetic penetration length $\lambda$, but coherence effects 
are also characterized by a coherence length $\xi$, with their ratio 
distinguishing between Type I and Type II superconductors. 
Namely, given the GL parameter $\kappa=\lambda/\xi$, 
Type I superconductors correspond to a value of the GL parameter
less than $1/\sqrt{2}$, $\kappa<\kappa_c=1/\sqrt{2}$,
and Type II superconductors to a value larger than $\kappa_c$, 
$\kappa>\kappa_c$. The manner in which magnetic fields may
or may not penetrate such materials in their bulk is different for
each Type. In particular, Type II materials sustain Abrikosov 
vortices,\cite{Abrikov} namely flux tubes carrying a unit value 
of the quantum of flux penetrating the material even in 
the superconducting state.

Furthermore, in the limit that any spatial dependence of the order parameter
may be ignored, the GL equations lead back to yet an older empirical
approach to superconductivity from 1935 due to the London brothers.\cite{London}
The London equations simply read
\begin{equation}
\vec{E}=\frac{\partial}{\partial t}\left(\Lambda\vec{J}\right),\qquad
\vec{B}=-\vec{\nabla}\times\left(\Lambda\vec{J}\right),
\end{equation}
$\vec{E}$ and $\vec{B}$ being the electric and magnetic fields, respectively,
$\vec{J}$ the current density, and $\Lambda$ a phenomenological parameter
given by
\begin{equation}
\Lambda=\frac{m}{n_s q^2},
\end{equation}
$n_s$ being the density of superconducting electrons. Perfect conductivity
is a direct consequence of the first London equation, while the Meissner
effect follows from the second with a magnetic penetration length $\lambda_L$
such that
\begin{equation}
\lambda_L^2=\frac{m}{\mu_0 n_s q^2}.
\end{equation}

Of course, it is to be understood that all the above descriptions
and their ensuing equations of motion are also coupled to
Maxwell's equations of electromagnetism,
\begin{equation}
\begin{array}{rclrcl}
\vec{\nabla}\cdot\vec{E}&=&\frac{1}{\epsilon_0}\,\rho,\qquad &
\vec{\nabla}\times\vec{E}+\frac{\partial\vec{B}}{\partial t}&=&\vec{0},\\
 & & & & & \\
\vec{\nabla}\cdot\vec{B}&=&0,\qquad &
\vec{\nabla}\times\vec{B}-\frac{1}{c^2}\frac{\partial\vec{E}}{\partial t}
&=&\mu_0\vec{J},
\end{array}
\end{equation}
$\rho$ being the electric charge density, and $\epsilon_0$ et $\mu_0$
the usual electric and magnetic permittivity properties of the vacuum
such that $\epsilon_0\mu_0=c^2$, $c$ being the speed of light in vacuum.

But one of the players in these latter equations which is conspiscuously
missing from the above discussion of available descriptions of
LTc superconductivity is the electric field. How do electric fields influence or
affect the electromagnetic properties of superconducting materials?
The usual answer\cite{Tinkham,Waldram} to this question is that, 
in the stationary state without any time dependence, electric fields 
cannot have any effect whatsoever since, according to the first London 
equation and because of the perfect conductivity of any superconductor, 
the electric field must vanish identically at least for bulk materials. 
Given that this answer is presumably acceptable, it then remains nonetheless
possible that for nanoscopic materials of increasing use and interest
in nanotechnology, electric fields may have some effect close
to the surface of such materials, since it would be difficult to
imagine how an externally applied electric field could abruptly and
discontinuously vanish when moving from the outside to the inside 
of such a conductor.\cite{Gov1} In the present contribution, we briefly discuss
this question, and present some of the conclusions that have been
reached through the work of which far more details may be found in
Ref.~\refcite{These}.

The characterisation of the problem is presented in
Sec.~\ref{Sec2}. Next, a first framework in which to address the
issue is briefly considered in Sec.~\ref{Sec3}, with experimental
results proving that the analysis must be extended to include
the effects of all electrons of a superconducting material, even
the ``normal" ones. An appropriate framework is then developed
in Sec.~\ref{Sec4}, leading first to the identification of the
effective potential energy in analogy with the GL potential,
and next, in Sec.~\ref{Sec5}, some further dynamical properties
of the superconducting state in the presence of an applied
electric field. Finally, Sec.~\ref{Sec6} offers some conclusions
and prospects for further work along similar lines.

\section{The Problem}
\label{Sec2}

To highlight the issue mentioned above from different points of view,
let us first recall that the relativistic covariance properties
of Maxwell's equations are best made manifest through the fact that
the electromagnetic scalar and vector potentials, $\Phi$ and $\vec{A}$,
respectively, define the components of a 4-vector as
\begin{equation}
A^\mu=\left(\frac{\Phi}{c},\vec{A}\right),\qquad \mu=0,1,2,3,
\end{equation}
while the associated field strength tensor
$F_{\mu\nu}=\partial_\mu A_\nu-\partial_\nu A_\mu$ (with $x^\mu=(ct,\vec{x}\,)$)
is directly related to the electric and magnetic fields $\vec{E}/c$ and $\vec{B}$ as
\begin{equation}
\frac{\vec{E}}{c}=-\vec{\nabla}\frac{\Phi}{c}-
\frac{\partial}{\partial(ct)}\vec{A},\qquad
\vec{B}=\vec{\nabla}\times\vec{A}.
\end{equation}
The fact of the matter is that all the approaches briefly reviewed
in the Introduction are intrinsically nonrelativistic, but are
nonetheless coupled to the relativistic covariant
Maxwell's equations. In itself this is not necessarily problematic
provided the considered regime remains nonrelativistic in a sense
to be specified. However, since a nonrelativistic limit amounts to
taking a limit such that $1/c\rightarrow 0$, clearly any of the
effects related to the electric scalar potential, $\Phi/c$, and field,
$\vec{E}/c$, in the same units as those of the magnetic sector,
decouple in such a limit. Electric field effects in superconductors
are thus at best subleading in $1/c$, but not necessarily vanishing
altogether. Therefore one ought to develop a manifestly relativistic
invariant framework in which to analyse such effects.

From yet another point of view, one may also argue for the necessity
of such a framework by considering specific experimental set-ups.\cite{Gov1}
For example, imagine an infinite slab subjected to an external magnetic
field parallel to it, without an external electric field
being applied in the laboratory frame. Due to the Meissner effect,
the magnetic field will penetrate the slab only up a typical distance
set by the magnetic penetration length $\lambda$. Imagine now performing 
a Lorentz boost in a direction both parallel to the slab and perpendicular
to the magnetic field. In such a boosted frame, not only is the
strength of the magnetic field slightly modified, but more importantly
there appears now an electric field perpendicular to the slab,
and {\it a priori\/} also inside the superconductor, and thus necessarily
with precisely the same penetration length as the magnetic field!
Hence, if for such a {\sl gedanken experiment\/} it is justified to
restrict only to electrons in the superconducting state, one is forced
to conclude, on basis of relativistic covariance, that an electric
field does not necessarily vanish inside a superconductor, and does
penetrate such materials with a penetration length identical to the
magnetic one.

Such considerations thus call for a framework in which Maxwell's
electromagnetism is coupled to the superconducting state in a
manifestly relativistic invariant manner. Such a framework may also
be of relevance to other issues of superconductivity, especially
for heavy metallic coumpounds corresponding to chemical elements
of large $Z$ values, implying significant relativistic corrections
to electronic orbital properties.\cite{Capelle,Ohsaku}

In effect, an experiment such as the one described above has been
performed using a nanoscopic slab of superconducting aluminium,
subjected to both a magnetic field parallel to the slab and an electric
field perpendicular to it (for the details, see Ref.~\refcite{These}).
Much to our surprise, no effect of the electric field whasoever, 
even when ramped up to considerable values, was observed on the critical 
temperature for the superconducting state, while the latter's values 
and dependence on the magnetic field were properly observed, measured 
and seen to coincide with established values for the critical temperature 
in the case of that material.  

The experiment was analysed within the framework of both the nonrelativistic
GL approach, and its obvious covariant generalisation through the
U(1) Higgs model for a charged complex scalar field with Lagrangian
density functional\cite{Weinberg,Gov1,Gov2}
\begin{equation}
{\cal L}=-\frac{1}{4}\epsilon_0 c F^{\mu\nu}F_{\mu\nu}\,+\,
\frac{1}{2}\epsilon_0c\left(\frac{\hbar}{q\lambda}\right)^2
\left\{\left|\left(\partial_\mu+i\frac{q}{\hbar}A_\mu\right)\psi\right|^2
-\frac{1}{2\xi^2}\left(|\psi|^2-1\right)^2\right\}.
\end{equation}
In the latter case, indeed the London equations are modified in the
manner expected on account of manifest Lorentz covariance,
with in particular necessarily identical electric and magnetic
penetration lengths.\cite{Gov1}

Either framework predicts effects of an electric field on such
an experimental set-up, with specific characterisations of these
effects as a function of the temperature.\cite{Gov1} And much to our surprise,
no effect whasoever was observed in spite of repeated and carefully
prepared measurements and nanoscopic samples of the aluminium slab.\cite{These}

Faced with this conundrum, we were led to the necessity of developing
a microscopic model of $s$-wave superconductivity which is manifestly
relativistic invariant and accounts for all electronic states,
whether ``superconducting" or ``normal" states, the latter being the main
suspects as being behind the close-to-perfect screening of
any elecric field however large. In other words, a relativistic extension
of the BCS theory for LTc superconductors is {\it a priori\/} required
to account for the experimental results.

\section{A Model}
\label{Sec3}

The superconducting state we are interested in being in thermodynamical 
equilibrium, the natural framework to model the problem at 
the microscopic level is in terms of Finite Temperature Quantum Field 
Theory (FTQFT),\cite{Kaputsa,LeBellac} in which electron states 
are described by the Dirac field coupled to the electromagnetic field 
in a U(1) gauge invariant manner. One needs to compute the partition 
function of the system, in the presence of stationary background 
electric and magnetic fields, as well as a chemical 
potential\cite{LeBellac,Feynman} $\mu$ for the electron states, namely
\begin{equation}
Z(\beta)={\rm Tr}\,e^{-\beta\left(H-\mu N\right)},\qquad
\beta=\frac{1}{kT},
\end{equation}
$H$ being the Hamiltonian of the system, $N$ its electron number operator,
$T$ the absolute temperature and $k$ Boltzmann's constant, while the
trace is over all quantum states of the system. The calculation
of such a partition function proceeds through both operatorial and
path integral techniques.

The Hamiltonian to be used stems from the Lagangian density describing
the microscopic dynamics
\begin{equation}
{\cal L}=-\frac{1}{4}F_{\mu\nu}F^{\mu\nu}+
\overline{\psi}\left(i\gamma^\mu\partial_\mu-m\right)\psi
-eA_\mu\overline{\psi}\gamma^\mu\psi+
{\cal L}_{\rm 4f}.
\end{equation}
Henceforth, natural units such that $\hbar=1=c$ and $\epsilon_0=1=\mu_0$ 
are used. Here, $\psi$ stands for a Dirac 4-spinor for the Dirac-Clifford
algebra $\{\gamma^\mu,\gamma^\nu\}=2\eta^{\mu\nu}$, $\eta^{\mu\nu}$
being the four-dimensional Minkowski spacetime metric of mostly negative
signature (for the space components), $m$ is the electron mass
and $e<0$ its electric charge, while $\overline{\psi}=\psi^\dagger\gamma^0$. 
Finally, ${\cal L}_{\rm 4f}$ stands for the four-fermion effective interaction 
to be used to model the phonon-mediated electron-electron interaction 
responsible for the superconducting state.

{\it A priori}, the 4-fermion interaction may involve some combination
of all possible relativistic invariant 4-electron operators, of the
form,
\begin{equation}
{\cal L}_{\rm 4f}=g_1\left(\overline{\psi}\psi\right)^2 +
g_2\left(\overline{\psi}\gamma_5\psi\right)^2 +
g_3\left(\overline{\psi}\gamma^\mu\psi\right)^2 +
g_4\left(\overline{\psi}\gamma^\mu\gamma_5\psi\right)^2 +
g_5\left(\overline{\psi}\sigma^{\mu\nu}\gamma_5\psi\right)^2,
\end{equation}
with, as usual, $\gamma_5=i\gamma^0\gamma^1\gamma^2\gamma^3$ and
$\sigma^{\mu\nu}=i[\gamma^\mu,\gamma^\nu]/2$, and $g_i$ ($i=1,2,3,4,5$)
arbitrary real couplings constants. However, since the model describes
a single fermionic species of which the field degrees of freedom
are represented by Grassmann odd variables, and applying Fierz
identities, it follows that the same effective interaction may be
brought into the form
\begin{equation}
\begin{array}{rcl}
{\cal L}_{\rm 4f}&=&\beta_1\left(\overline{\psi_c}\psi\right)^\dagger
\left(\overline{\psi_c}\psi\right) +
\beta_2\left(\overline{\psi_c}\gamma_5\psi\right)^\dagger
\left(\overline{\psi_c}\gamma_5\psi\right) \\
 & & \\
& & + \beta_3\left(\overline{\psi_c}\gamma^\mu\gamma_5\psi\right)^\dagger
\left(\overline{\psi_c}\gamma_\mu\gamma_5\psi\right),
\end{array}
\end{equation}
where $\psi_c=\eta_c C\overline{\psi}^T$, $\eta_c$ being an arbitrary
phase factor and $C$ the charge conjugation matrix operator. A detailed
analysis of the particle and spin content of these different operators
in a nonrelativistic limit shows that, in the order in which they appear
in the above relation, the first accounts for a $p$-wave order parameter,
the second the $s$-wave BCS state, and the third a superposition
of $d$- and $s$-wave contributions. This classification does not
coincide with different conclusions available in the literature,
which failed to account for the Grassmann character of the electron field
and misidentified the proper properties of such a classification under
parity.\cite{Capelle,Ohsaku}

Focusing on $s$-wave BCS superconductivity, we are thus led to consider
the following 4-fermion effective interaction
\begin{equation}
{\cal L}^{BCS}_{\rm 4f}=-\frac{1}{2}g\,
\left(\overline{\psi_c}\gamma_5\psi\right)^\dagger
\left(\overline{\psi_c}\gamma_5\psi\right),
\label{eq:order}
\end{equation}
$g$ being a real coupling constant, in order to model the phonon mediated
interaction between electron pairs and leading {\it in fine\/}
to the Cooper pair condensed state below the critical temperature.

Given that choice as well as well established techniques 
of FTQFT,\cite{LeBellac} it follows that the partition function 
of the system may be given the following path integral representation,
\begin{equation}
Z(\beta)=\int{\cal D}[\psi,\overline{\psi},\Delta,\Delta^\dagger]\,
e^{-\int_0^\beta d\tau\int d^3\vec{x}\,{\cal L}_{\rm E}},
\end{equation}
where
\begin{equation}
\begin{array}{rcl}
{\cal L}_{\rm E}&=&
\frac{1}{2}\left[\psi^\dagger\partial_\tau\psi\,-\,
\partial_\tau\psi^\dagger\psi\right]\,-\,
\frac{1}{2}i\left[\overline{\psi}\vec{\gamma}\cdot\vec{\nabla}\psi\,-\,
\vec{\nabla}\overline{\psi}\cdot\vec{\gamma}\psi\right]\\
 & & \\
& & + m\overline{\psi}\psi + e A_\mu\overline{\psi}\gamma^\mu\psi
-\mu_0\psi^\dagger\psi\\
 & & \\
& & +\frac{1}{2g}|\Delta|^2-\frac{1}{2}\left[
\Delta^\dagger\left(\overline{\psi_c}\gamma_5\psi\right)\,+\,
\Delta\left(\overline{\psi_c}\gamma_5\psi\right)^\dagger\right],
\end{array}
\end{equation}
while from now on $\mu_0$ stands for the chemical potential in the absence
of external electromagnetic fields.
Here, $\tau$ is the imaginary time parameter in which bosonic fields
must be periodic and fermionic ones antiperiodic with period $\beta$,
while $\Delta$ is an auxiliary field which is introduced in order
to express the 4-fermion interaction in terms of only quadratic couplings
of the Dirac field. The ensuing Grassmann odd gaussian integrals
are then readily feasable, leading to an effective action for the
auxiliary field $\Delta$. As a matter of fact, the field $\Delta$
coincides thus with the order parameter of the superconducting
state, and measures, up to normalisation, the local density of
Cooper states in that state. The effective action obtained through
the integration over all fermionic degrees of freedom thus
corresponds, in the relativistic setting, to the GL action functional 
in the nonrelativistic setting. In effect, this is also how Gor'kov 
established the GL effective description from
the BCS microscopic one.\cite{Gorkov}

\section{The Effective Potential}
\label{Sec4}

Before addressing the issue of the external field dependence of the
effective action, it is of interest to identify the effective
potential independently of such external electromagnetic disturbances
and spatial variations of the order parameter. For all practical purposes,
this effective potential should correspond to the GL potential
of the Higgs type in (\ref{eq:Higgs}).

In such specific circumstances, given the absence of external
electromagnetic fields and the assumption of a space independent
order parameter $\Delta_0$, the calculation may be performed
exactly through operator techniques by relying on a Bogoliubov
transformation which enables one to identify the associated
Cooper pairs in analogy with (\ref{eq:Cooper}). Details
may be found in Ref.~\refcite{These}. What such a Bogoluibov transformation
achieves is an exact diagonalisation of the quantum Hamiltonian
under the above circumstances, with the Cooper pair condensate defining
the physical ground state. Furthermore, excitations of the Cooper pair
condensate correspond to collective modes of the electron system of
definite momentum and charge, hence also of definite energy, known as pseudo-particles. 
Any of these states, whether the Cooper pair ground state or its
pseudo-particle excitations, are obtained as coherent superpositions
of the modes of the original free quantum electronic states of the
Dirac spinor and its perturbative vacuum. 

Denoting by $V$ the volume of the superconductor, the effective 
potential energy density is expressed as
\begin{equation}
\begin{array}{rcl}
\frac{1}{V}S^{(0)}_{\rm eff}&=&
\frac{1}{2g}|\Delta_0|^2
+\int\frac{d^3\vec{k}}{(2\pi)^3}
\left[2\omega(\vec{k}\,)-E_B(\vec{k}\,)-E_D(\vec{k}\,)\right]\\
 & & \\
& & -2\frac{1}{\beta}\int\frac{d^3\vec{k}}{(2\pi)^3}\,
\ln\left[1+e^{-\beta E_B(\vec{k}\,)}\right]\,-\,
2\frac{1}{\beta}\int\frac{d^3\vec{k}}{(2\pi)^3}\,
\ln\left[1+e^{-\beta E_D(\vec{k}\,)}\right],
\end{array}
\end{equation}
where
\begin{equation}
\begin{array}{rcl}
E_B(\vec{k}\,)&=&\sqrt{\left(\omega(\vec{k}\,)-\mu_0\right)^2+|\Delta_0|^2},\\
 & & \\
E_D(\vec{k}\,)&=&\sqrt{\left(\omega(\vec{k}\,)+\mu_0\right)^2+|\Delta_0|^2},\\
 & & \\
\omega(\vec{k}\,)&=&\sqrt{\vec{k}\,^2+m^2}.
\end{array}
\label{eq:Veff}
\end{equation}
Here, $E_B(\vec{k}\,)$ and $E_D(\vec{k}\,)$ stand for the energies
of the pseudo-particle excitations of electron and positron type,
respectively, in presence of the Cooper pair condensate $\Delta_0$
which clearly specifies also the gap value in the dispersion relations
for such collective excitations of the superconducting state.
Note that the chemical potential $\mu_0$, in the case of an electron
conductor,\footnote{In the case of a positron conductor, $\mu_0$
would be bounded above by $-m$.} is bounded below by $m$, since
the electron's relativistic rest mass energy must be added to the
ordinary nonrelativistic chemical potential value in the present Lorentz
covariant framework.

By minimisation of the effective potential and applying the usual
BCS approximation which consists in restricting the momentum integration
to a region surrounding the Fermi level with a cut-off set to coincide
with the Debye lattice frequency to which a Debye energy $\xi_D$ is
associated,\cite{BCS,Tinkham,Waldram} one obtains the following 
gap equation for the order parameter $\Delta_0$,
\begin{equation}
\frac{1}{2}\int_{-\xi_D}^{\xi_D}\,d\xi\,
\left\{\frac{1}{\sqrt{\xi^2+|\Delta_0|^2}}\,
\tanh\frac{1}{2}\beta\sqrt{\xi^2+|\Delta_0|^2}\right\}=
\frac{1}{gN(0)},
\label{eq:gap}
\end{equation}
$N(0)$ being the density of states at the Fermi level. In this
expression, possible contributions due to positron-like
states are not retained since their value is totally insignificant
in the case of an ordinary LTc superconductor. This gap
equation may be seen to coincide with the usual BCS gap equation.\cite{BCS}
In particular, in the weak coupling regime and at $T=0$ K, its solution reads
\begin{equation}
|\Delta_0(0)|\simeq\frac{\pi e^{-\gamma}}{\beta_c}\,
\frac{1}{1-e^{2/gN(0)}},
\end{equation}
$\gamma$ being the Euler constant, $\gamma\simeq 0.577$, and $\beta_c=1/kT_c$.
Hence, one recovers the BCS results, and the order parameter interaction
chosen in (\ref{eq:order}) does indeed represent $s$-wave BCS Cooper
pairs within the present relativistic framework. Given the kinematical
regime in which the model is being considered, relativistic corrections
to the effective potential thus prove to be totally insignificant.

Even though we shall refrain by lack of space
from presenting here graphs of the effective
potential (which would be quite illustrative of the physical
results, for which the interested reader is again referred to
Ref.~\refcite{These}), the effective potential (\ref{eq:Veff}) is
indeed of the Higgs type below the critical temperature, namely
with an absolute minimum for a nonvanishing order parameter
$\Delta$. However, although a quartic approximation of the Higgs form
\begin{equation}
V^{(1)}_{\rm eff}(x)=F(x_0)+
\left[F(0)-F(x_0)\right]
\left[\frac{x^2}{x^2_0}-1\right]^2, \qquad x=|\Delta|,
\end{equation}
is quite satisfactory for temperatures $T$ sufficiently close to $T_c$
and $\Delta$ values sufficiently close to the solution to the gap
equation (\ref{eq:gap}), and in which the coefficients $F(0)$ and $F(x_0)$
are chosen to coincide with the values of $V^{(1)}_{\rm eff}(x)$ for $x=0$ and
$x=x_0$, $x_0=|\Delta_0|$ standing for the solution to the gap equation,
better approximations are possible, which greatly extend the temperature
range below $T_c$ for which for all practical purposes the
approximation coincides with the exact effective potential given by
its integral definition in (\ref{eq:Veff}). Possible examples are\cite{These}
\begin{equation}
V^{(2)}_{\rm eff}(x)=F(x_0)+\left[F(0)-F(x_0)\right]
\left[\frac{\ln(x^2_0+\lambda x^2_0)-\ln(x^2+\lambda x^2_0)}
{\ln(x^2_0+\lambda x^2_0)-\ln(\lambda x^2_0)}\right]^2,
\end{equation}
\begin{equation}
V^{(3)}_{\rm eff}(x)=F(x_0)+\left[F(0)-F(x_0)\right]
\left[\,1-\left(\frac{\left[1+\gamma\frac{x^2}{x^2_0}\right]^\alpha-1}
{\left[1+\gamma\right]^\alpha-1}\right)^2\,\right],
\end{equation}
where $\lambda$, $\alpha$ and $\gamma$ are parameters whose values and
temperature dependence may be fitted\cite{These} to the exact effective potential,
leading to very efficient approximations to the exact expression,
reliable in far greater temperature and order parameter
ranges away from their critical values
than the usual Higgs potential of the quartic type, $V^{(1)}_{\rm eff}(x)$.
A study of the phenomenological consequences of such generalised
Higgs-like potentials could be of interest, in particular for what concerns
their vortex solutions.\cite{Gov2,Gov3,Stenuit}

\section{The Effective Action}
\label{Sec5}

By including the effects of external electromagnetic fields,
a computation of the full effective action, and not only the
effective potential, is feasible through a perturbative expansion.
Namely by also including effects due to space gradients both
in the order parameter and the electromagnetic potentials $\Phi$ and $\vec{A}$,
it is possible to obtain explicit expressions for all physically
relevant parameters which empirically characterise the superconducting
state. Thus not only are the coherence and magnetic penetration
length values and their temperature dependences obtained, but also
those of the electric penetration length.
Furthermore, other characterisations also become accessible,
which are usually not discussed in the literature by lack of
interest in the possible effects of electric fields on superconductors.
For instance, it is also possible to study how the total electron
charge density distribution, which ought to balance the background lattice
charge distribution, is accounted for by the order parameter
(``superconducting electrons"), and thus how superconductors could
locally acquire charge in specific circumstances. Likewise, a study of
the dependence of all the above quantities on the chemical potential 
$\mu_0$ is also feasible, and is of interest since varying its value
amounts to depleting the superconductor of its electrons, or else increasing
that number, in other words, charging the material.

By lack of space, all the results obtained so far along such lines
are not detailed here. They are available in Ref.~\refcite{These}
together with relevant and illustrative graphs. In the case study of aluminium,
experimental values for the magnetic penetration length and its 
temperature dependence are well reproduced from our analysis
when proper account is given of the role of impurity electron rescattering.
For what concerns the electric penetration length, our analysis
reveals that this observable, heretofore never computed in the
literature, receives two types of contributions, in contradistinction
to the magnetic penetration length of which the value is solely dependent
on the Cooper pair condensate density $|\Delta_0|$. Indeed, for
the electric penetration length, not only is there a contribution akin
to the magnetic penetration length as expected by reason of
the Lorentz covariance arguments discussed in Sec.~\ref{Sec2}, but in
addition the ordinary Thomas--Fermi screening effect existing in
normal conductors\cite{Kittel} is also at work. Since typically values 
for the latter screening length are on the order of the Angstr\"om or 
in fact even less, while magnetic penetration lengths typically range 
in the tens to hundreds of nanometers, and since their combined effect 
which finally sets the electric penetration length derives essentially 
from the sum of their squared inverse values, namely
\begin{equation}
\lambda^{-2}_{\rm electric}=
\lambda^{-2}_{\rm magnetic}+\lambda^{-2}_{\rm Thomas-Fermi},
\end{equation}
it follows that it is the Thomas--Fermi screening effect which by far
and large dominates the screening effects of electric fields in
superconductors. This conclusion also explains the null results
of our experimental measurements mentioned in Sec.~\ref{Sec2}.
In other words, Cooper pair contributions, namely ``superconducting
electron" contributions are indeed similar in value for both the magnetic and 
electric penetration lengths, but in the latter case contributions from
``normal electrons" are also involved, and their effect being so
overwhelming in the case of that observable, in effect
the complete superconducting electric penetration length
essentially coincides with the Thomas--Fermi screening length
of the conductor even in the normal conducting state.

In fact, since all electron states are being integrated out
in the path integral leading to the effective action, the
distinction between ``superconducting" and ``normal"
electrons is a matter of convention and arbitrary definition,
possible for instance by comparing local charge distributions
to the background lattice charge distribution which is also,
up to the sign, that of the conducting electrons in the normal state.

In technical terms, the electric penetration length is identified
directly from the effective potential rather that the full
effective action, through the dependence of the effective potential
(\ref{eq:Veff}) on the chemical potential. Indeed, the chemical
potential adds up with the electrostatic potential, and the
actual effective action is then function of the electrochemical
potential. Through an expansion in the electrostatic potential,
one then identifies the electric penetration length. More specifically,
given the effective action computed as indicated above through the
path integral over the fermionic degrees of freedom, one still needs
to add to it the Hamiltonian or energy density of the purely
electromagnetic sector, which is treated semi-classically in the
effective field theory approach. The latter reads
\begin{equation}
\begin{array}{rl}
& \int d^3\vec{r}\,\left\{\frac{1}{2}\vec{E}^2+\frac{1}{2}\vec{B}^2-
\Phi\left(\vec{\nabla}\cdot\vec{E}-\rho_{\rm tot}\right)\right\} \\
 & \\
= & \int d^3\vec{r}\,\left\{
\frac{1}{2}\left(\vec{E}+\vec{\nabla}\Phi\right)^2+\frac{1}{2}\vec{B}^2
-\frac{1}{2}\left(\vec{\nabla}\Phi\right)^2+\Phi\rho_{\rm tot}\right\},
\end{array}
\end{equation}
where $\rho_{\rm tot}$ stands for the total charge density in the
conductor, inclusive of the background lattice and valence electron
contributions (the ``static" charges), to which those of conducting 
electrons modeled through the above discussion are to be added.
Consequently, for what concerns a stationary configuration,
by adding this contribution to that following from the effective action
one is left with a local functional of the form
\begin{equation}
{\cal L}_{\rm tot}=f\,\left|\left(\vec{\nabla}-2ie\vec{A}\right)\Delta\right|^2+
\Phi\rho_{\rm tot}-\frac{1}{2}g\Phi^2
-\frac{1}{2}\left(\vec{\nabla}\Phi\right)^2+
\frac{1}{2}\left(\vec{\nabla}\times\vec{A}\right)^2,
\end{equation}
where $f$ and $g$ are quite involved expressions determined from the
effective action calculation. Note well however that the term
in $g\Phi^2$ derives solely from the effective {\sl potential\/}
rather than the full effective action, whereas the term involving $f$
is a contribution from the effective action {\it per se\/} which
determines the magnetic penetration length. Deriving
now the equation of motion for the electrostatic potential,
\begin{equation}
\vec{\nabla}^2\Phi-g\,\Phi+\rho_{\rm tot}=0,
\end{equation}
it is quite clear that the electric field penetration length is
determined by the coefficient $g$ through
\begin{equation}
\frac{1}{\lambda^2_{\rm electric}}=g.
\end{equation}
Hence indeed the electric field penetration length is solely determined
from the expansion to second order of the effective potential
with respect to the chemical potential, since in the presence
of an external electrostatic potential the effective potential
is function of the electrochemical potential, namely the sum
of both the chemical and electrostatic potentials. Finally,
the explicit analysis finds that $g$ does receive two types of
contributions, one which vanishes for a vanishing order parameter
$\Delta_0$, {\it i.e.\/}, in the absence of Cooper pairs or
the ``superconducting electron" contribution, and the second
which remains finite even when $\Delta_0=0$, namely the ``normal electron"
contribution, which for all practical purposes leads in effect
to the Thomas--Fermi length.

Given this fact, it thus appears that a similar calculation is perfectly 
feasible also in a nonrelativistic setting, since one only needs to consider
the dependence of the effective potential on the chemical potential.
Nevertheless, and somewhat suprisingly perhaps, this dependence
does not appear to ever have been studied previously, and our
result is thus totally new in the literature.

Given that the Thomas--Fermi screening length increases with the
depletion of conducting electrons, {\it i.e.\/}, by charging positively
the conductor, it would appear that possibly one could in effect
remove the effect of the ``normal" electrons and thus reach a
regime in which both the magnetic and electric penetration lengths
have comparable values, enabling an experimental confirmation of
the effects of electric fields in a set-up of the type used in
our experiments. Unfortunately, a detailed analysis of the
dependence on the chemical potential $\mu_0$ of both these
penetration lengths given our explicit resuls, has established
that such a regime is never achieved. Even though both lengths 
essentially diverge when the conductor is totally depleted of
its conducting electrons, their ratio never approaches a value
close to unity, rather it essentially retains its value for the neutral
conductor. Likewise for what concerns their dependence on temperature,
although the magnetic penetration length diverges close to $T_c$,
the electric one does not display any particular behaviour when
crossing the critical threshold because of the dominance of the
Thomas--Fermi screening length, and one reproduces the correct
values and temperature dependence of the latter quantity in the normal
state as well.

Finally, our analysis has provided for the first time the temperature
dependence of the coherence length of a superconductor such as aluminium.
Even though the numerical values obtained for that material
coincide with measured ones, our explicit expressions for that
quantity lead to an unexpected behaviour of that observable when
approaching the critical temperature. Indeed, while it remains
rather stable at low temperatures, upon approaching $T_c$, one
observes first a slight dip (on the order to 10\%) in its value before
increasing as expected phenomenologically within the GL framework
as $T_c$ is reached. Note that in contradistinction to the magnetic
penetration length which is accessible experimentally, there does
not appear to exist measurements available in the literature of
the temperature dependence of the coherence length of superconductors.
This unexpected behaviour of the coherence length requires further
corroboration, both experimental and theoretical.

\section{Conclusions and Prospects}
\label{Sec6}

In this brief contribution, we have described some of the results
achieved through a relativistic invariant extension of the well
established BCS theory of LTc superconductivity. The motivation for
this study is a better understanding, in a relativistic regime
at a later dynamical stage, of the effects not only of magnetic fields
but also of electric fields on the superconducting state. Following
experimental measurements performed on nanoscopic superconductors
of which the results were totally unexpected, it was realised that
the role of ``normal" electrons is also crucial for what concerns
such electric field effects. When these are properly included within
a microscopic framework, it appears that ordinary screening effects
in conductors overwhelm the properties stemming from the
superconducting state, an occurrence which does not apply to
magnetic field effects for which only the contributions
from ``superconducting" electrons are relevant. Having identified
precisely the origin of the different contributions to the total
electric penetration length, it appears that, at least in the
instance of that specific observable, a nonrelativistic analysis of
the effective potential and its dependence on the chemical potential
would have sufficed to reach the same conclusion. In the regime
of temperatures and chemical potentials of relevance to ordinary
superconductors, indeed the effect of positron-like states is
perfectly insignificant. There exist other physical environments
though, for which this would no longer be the case, for instance in
the astrophysical context.

Our work leaves open a series of issues and even possibilities
of detailed study which deserve to be investigated further. For instance,
our analysis predicts that under certain circumstances
superconductors would acquire locally on their surface nonvanishing
charge. Since in recent years such measurements have become possible,
a detailed analysis of this issue would be of interest with the
prospect of experimental validation of the model. An unexpected
temperature dependence of the coherence length has been identified,
which deserves confirmation. The relativistic framework has also
led to further possible types of order parameters than simply the
$s$-wave BCS one, including $p$- and $d$-wave order parameters.
As is well known, High Temperature (HTc) superconductors display
properties of mixtures of $s$-, $p$ and $d$-wave order parameters,
in combinations depending on the material being considered.
Would the present framework enable a description of some of
these HTc superconductors? Note also that by combining now
in the effective four-fermion interaction a superposition of the
three types of order parameters, one would obtain a description
of systems possessing more than one gap, indeed as also been
observed for some HTc materials. There is thus a rich phenomenology
of properties to be described through such generalisations of our
work. Finally, one may still extend further the choice of
four-fermion interaction, by including higher derivative couplings
or introducing Lorentz noninvariant couplings. Indeed after all,
the thermodynamical description remains tied up with the rest frame of the
material being considered, and from that point of view one may
still extend the range of possible four-fermion interactions,
and see whether some classes of models could account for the
observed properties of new classes of superconductors.

\section*{Acknowledgements}

J. G. acknowledges the Abdus Salam International Centre for Theoretical
Physics (ICTP, Trieste, Italy) Visiting Scholar Programme 
in support of a Visiting Professorship at the International Chair 
in Mathematical Physics and Applications (ICMPA). This work
is partially supported by the Belgian Federal Office for Scientific, 
Technical and Cultural Affairs through the Interuniversity Attraction 
Pole (IAP) P5/27.

\end{document}